\documentclass[sigconf]{acmart}

\AtBeginDocument{%
  \providecommand\BibTeX{{%
    \normalfont B\kern-0.5em{\scshape i\kern-0.25em b}\kern-0.8em\TeX}}}

\setcopyright{acmcopyright}
\copyrightyear{2021}
\acmYear{2021}
\acmDOI{10.1145/1122445.1122456}

\newcommand{\excessin}{\mathcal{I}_E}

\newcommand{\timex}{t}

\newcommand{\businter}{B_{\Delta t}}
\newcommand{\busintersched}{B_{\Delta t, sched}}
\newcommand{\lastthree}{s_{p,3}}
\usepackage{color}
\usepackage{caption}
\usepackage{subcaption}

\newtheorem{definition}{Definition}

\begin{document}

\title{Excess demand in public transportation systems: The case of Pittsburgh's Port Authority}

\author{Tianfang Ma}
\affiliation{%
  \institution{Graduate School of Public and International Affairs, University of Pittsburgh}
  \city{Pittsburgh}
  \state{PA}
  \country{USA}
  \postcode{15260}
}

\author{Robizon Khubulashvili}
\affiliation{%
  \institution{Graduate School of Public and International Affairs, University of Pittsburgh}
  \city{Pittsburgh}
  \state{PA}
  \country{USA}
  \postcode{15260}
}

\author{Sera Linardi}
\affiliation{%
  \institution{Graduate School of Public and International Affairs, University of Pittsburgh}
  \city{Pittsburgh}
  \state{PA}
  \country{USA}
  \postcode{15260}
}

\author{Konstantinos Pelechrinis}
\affiliation{%
  \institution{Department of Informatics and Networked Systems, University of Pittsburgh}
    \city{Pittsburgh}
  \state{PA}
  \country{USA}
  \postcode{15260}}

\renewcommand{\shortauthors}{Ma, et al.}

\begin{abstract}
  ``An advanced city is not a place where the poor move about in cars, rather it's where even the rich use public transportation''. 
This is what Enrique Pe\~{n}alosa, the celebrated ex-mayor of Bogota once said. 
However, in order to achieve this objective, one of the crucial properties that the public transportation systems need to satisfy is {\em reliability}. 
While reliability is often referenced with respect to on-schedule arrivals and departures, in this study we are interested in the ability of the system to satisfy the total passenger demand. 
This is crucial, since if the capacity of the system is not enough to satisfy all the passengers, then ridership will inevitably drop. 
However, quantifying this excess demand is not straightforward since public transit data, and in particular data from bus systems that we focus on in this study, only include information for people that got on the bus, and not those that were left behind at a stop due to a full bus. 
In this work, we design a framework for estimating this excess demand. 
Our framework includes a mechanism for identifying instances of potential excess demand, and a Poisson regression model for the demand for a given bus route and stop. 
These instances of potential excess demand are filtered out from the training phase of the Poisson regression. 
We show through simulated data that this filtering is able to remove the bias introduced by the censored data logged by the system. 
Failure to remove these data points leads to an underestimation of the excess demand. 
We then apply our approach on real data collected from the Pittsburgh Port Authority and estimate the excess demand over an one-year period. 
\end{abstract}



\maketitle

\section{Introduction}\label{sec:intro}

One of the crucial properties that the public transportation systems need to satisfy is that of {\em reliability}. 
While reliability is often referenced with respect to on-schedule arrivals and departures, in this study we are interested in a different dimension, namely, the ability of the system to satisfy the total passenger demand. 
This is crucial, since if the capacity of the system is not enough to satisfy all the passengers the ridership will inevitably drop through a negative feedback loop driven by the low quality service. 
Typically transit system operators have tried to provide solutions to this problem through congestion pricing. 
For example, in Singapore subway commuters traveling before the morning rush hours could get free or discounted rides depending on their route/destination. 
The program was able to reduce the commuters peak-off peak ratio from 2.8-to-1 to 2-to-1 \cite{singapore}. 
Other cities have had plans of replicating this approach \cite{toronto}. 
Smoothing out the peak-off peak ratio is certainly beneficial for both the transit agencies and the commuters, but given that many (possibly the majority) of commuters during rush hours might not have the time flexibility required to change their commute times, the problem of excess demand is still present (albeit to a lesser degree). 

While transit agencies are trying to minimize excess demand, quantifying the latter is not trivial and to the best of our knowledge there are not any approaches of doing so (at least in the public).  
Even the evaluation of the Singapore congestion pricing experiment, used as an evaluation metric the peak-off peak ratio of the number of commuters. 
This does not tell us anything about the number of commuters that were not able to board on the vehicle and had to wait for the next one or chose a different mode of transportation altogether. 
Quantifying this excess demand is not straightforward since public transit data - and in particular data from bus systems that is the focus of our study - only include information for people that got on the bus, and not those that were left behind at a stop due to a full bus.
Therefore an observation of 0 passengers boarding on a bus does not necessarily mean that no one wanted to get on the bus.

The objective of our work is to understand the characteristics of excess bus demand and design a framework for estimating it. 
We start by developing a mechanism to detect instances, that is, bus arrivals at stops, that might exhibit excess demand. 
This mechanism is essentially a binary classification, where the positive class corresponds to ``presence of excess demand'', while the negative class corresponds to ``no excess demand''. 
The recorded data for the (true) positive instances will be inevitably censored, since the logged number of passengers boarding the bus will not include those passengers left behind. 
Using simulated data we show that this identification is crucial, since if we train a passenger demand model including these censored data we will end up underestimating the excess demand. 
The degree of underestimation increases as the volume of the excess demand increases, while the improvement - as one might have expected - depends on how well the binary excess demand detection performs. 

Consequently, using data from the Pittsburgh Parking Authority and applying the aforementioned filtering we examine several models for count data for the passenger demand. 
In particular, we examine Poisson regression, zero-inflated Poisson regression, Negative Binomial regression and a hierarchical model. 
Our results indicate that the Poisson regression provides consistently the smallest error on our dataset among the other alternatives, even though the improvements might be marginal in some cases.  
With this model choice we estimate the excess demand in Pittsburgh's Port Authority system for the one year period covered by our data for the 10 most busy routes (as measured from the number of total passengers). 
Our estimations indicate that about 1\% of the total passengers of a route are left at the bus stop due to a full bus over the whole year, with this fraction exhibiting seasonality and being larger during the fall months. 
Furthermore, if we only focus on rush hours, the fraction of passengers left out is up to 8\% of the total passengers during that time period.



The rest of the paper is organized as follows. 
Section \ref{sec:related} discusses relevant literature to our study, while Section \ref{sec:Ie_detection} starts by describing the Pittsburgh Port Authority (PPA) data. 
We continue by describing in detail the setting(s) where excess demand appears and our mechanism for detecting these instances along with its limitations. 
We finally quantify through simulated data the benefits of removing the detected instances from training in terms of estimation bias. 
In Section \ref{sec:model_selection} we describe the model selection process on the real data from PPA, while we also quantify the excess demand in the system for the period covered in our data. 
Finally, Section \ref{sec:conclusion} concludes our study and also discussing its limitations. 
 
\section{Prior Literature}
\label{sec:related}

In this section we are going to discuss literature related to our study. 
We are also going to further differentiate our work.  

Studying the demand for public transportation has been a central task for transport engineers for several decades now. 
As (bus) transit systems were developed in the cities the operators wanted to understand and be able to predict the ridership demand of the system. 
One of the early studies to use data for modeling the demand for a bus system was that from Schmenner \cite{schmenner1976demand}. 
Schmenner built a regression model for the demand on a per route basis for 3 cities in Connecticut, namely, Hartford, New Haven and Stamford. 
Given the absence of detailed passenger data, he used the total revenue per route as the independent variable, since it is directly related to the ridership demand. 
He used a variety of controls for his model, including geographic and demographic information for the areas the route went through. 
Despite the longtime interest in estimating transit demand, a big obstacle for large scale studies was the unavailability of detailed passenger data. 
This led to studies focusing on smaller scale analyses on a specific area of a city, bus route and/or bus stop(s), theoretical modeling developments, or, macro-analysis of the interaction between policies and long-term ridership (e.g., \cite{doti1991model,bierlaire1997mathematical,ferguson1992transit,bussiere1984population,attaluri1997modeling} with the list not being exhaustive). 

However, the last few decades due to various technological advancements, such as, ``tap-in, tap-off'' ticketing systems, detailed and large-scale ridership data have been collected. 
This has consequently led to the development of ridership models at the bus stop level (e.g., \cite{kikuchi2001use,pulugurtha2012assessment,chu2004ridership,cervero2010direct}).  
The objective of these models is seemingly the same with ours, but there is a {\bf subtle difference that separates our study}. 
In particular, the independent variable for all the models in these studies is the number of passengers that boarded the bus at a given stop. 
However, this is not necessarily the ridership demand at the stop as we discussed in the previous section. 

Another line of work, tangential to these studies, deals with the estimation of the arrival time of a bus at a given stop. 
This is related with the excess demand, since a delayed bus will arrive to a bus stop with potentially more commuters waiting to board, hence, increasing the chances of excess demand. 
A variety of machine learning models have been used for this task ranging from support vector machines to neural networks (e.g., \cite{yu2011bus,bin2006bus,chien2002dynamic,shalaby2004prediction}). 
A smaller set of studies have also explored models for predicting/estimating the load on the bus (e.g., \cite{gayah2016estimating,arabghalizi2019full}). 
While modern systems keep track of the current load (i.e., the number of passengers) on the bus, the objective of these studies is to project the load of the bus when it reaches a specific bus stop.  

As it should be evident our work contributes to this literature by treating the observations of the number of passengers boarding on the bus at each stop as possibly censored. 
This will allow to obtain an estimate of the excess demand, i.e., the number of passengers left behind at each bus-stop/route combination. 
At this point we would like to mention that there are a few studies that have tried to explore the use of Internet of Things devices to estimate the number of people waiting at a bus stop and their wait time. 
For example, association of mobile devices with WiFi hotspots can be leveraged to obtain an estimate of the number of people waiting at a bus stop \cite{arai2021leveraging}.  
While these methods have shown promise, one of the major drawbacks is associated with the need for the appropriate infrastructure for this cyber-physical system to operate. 
Nevertheless, our work is complementary to these efforts, and it can actually benefit from small scale deployments that might be able to obtain actual ground truth for the excess demand. 

\section{Data and Excess Demand Instances Identification}\label{sec:Ie_detection}

In this section we will describe the data that we have obtained from PPA and we will use to estimate the excess demand for the system.
We also describe the setting under which excess demand appears, and a mechanism for detecting these instances. 
We further show through simulated data, where, unlike the PPA data, the ground truth of excess demand is known, that using these instances during the training of a passenger demand model leads to underestimation of the excess demand.

\subsection{PPA Data}\label{sec:data}

For our study we obtained data from Pittsburgh's Port Authority that includes detailed information for every bus route trip taken in the city between 1/1/18 and 12/31/18. 
The dataset covers 98 different routes and a total of 6102 bus stops.  
The number of trips per day per route ranges from 17 to 69.
For every trip recorded for a given route with $n$ stops, there are $n$ records; i.e., every record corresponds to a bus stop of the route. 
The information provided for each stop of the trip includes: 

\begin{itemize}
    \item {\tt ARRIVAL TIME}: this is the time the bus arrived at the bus stop
    \item {\tt ON}: this is the number of passengers that got on the bus at that stop
    \item {\tt OFF}: this is the number of passengers that got off the bus at that stop
    \item {\tt LOAD}: this is the total number of passengers on the bus when it leaves the stop
 \item {\tt CAPACITY}: this is the number of seats on the bus
\end{itemize}

This information is collected through infrared sensors (combined with GPS mounted on the bus for its location information) that are placed at the doors of the bus. 
This is an accurate and cost-effective solution for systems that do not have a unified ``tap-in'' payment infrastructure for fare collection. 
Therefore, the information at hand in terms of passenger load is aggregate, that is, we do not know where individual passengers get on and off the bus. 
PPA runs two types of buses to serve the routes, namely, single and articulated. 
Single buses can sit 40 people, while articulated can sit 56. 
These buses are running on an optimized schedule given the observations for the current ridership.  
For example, the articulated buses are deployed mostly in the 9am-6pm time slot, and never after 8pm. 
Port Authority bus drivers are also instructed to not allow any additional passenger to board on the bus if the bus has reached its {\bf capacity limit, which is considered to be equal to approximately 140\% of the number of seats}. 
This is when the excess demand will appear and passengers will not be able to board on the bus. 
Even though we have access to the bus load as described above, this information is not available to the bus driver in real time. 
Hence, it is very possible - almost certain - that there is noise associated with the decision drivers make to allow more passengers on board or not. 

\subsection{Excess Demand Detection Mechanism}
\label{sec:ed_det}

\begin{figure*}
     \centering
    \includegraphics[scale=0.55]{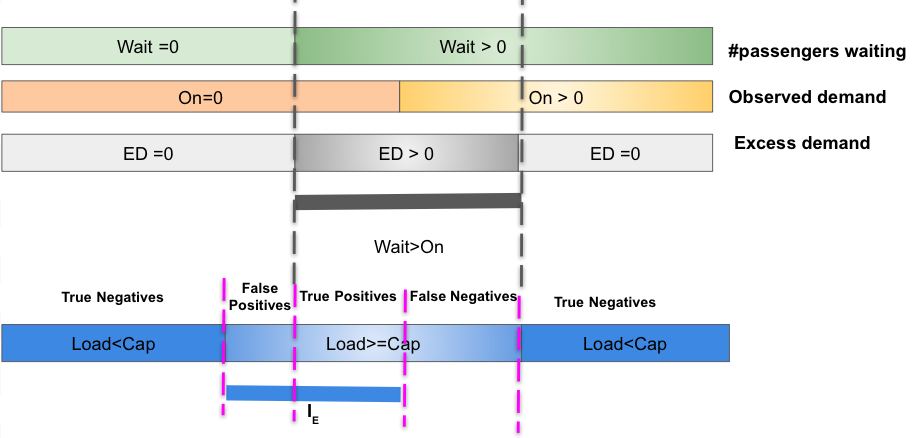}
    \caption{Different scenarios for the excess demand levels at a bus stop.}
    \label{fig:ed_viz}

\end{figure*}

Figure \ref{fig:ed_viz} depicts the possible settings that we have to consider at a bus stop during different arrivals. 
The first row represents the number of passengers that are waiting at the bus stop, while the second row corresponds to the number of passengers getting on. 
Finally, the third row corresponds to the excess demand, that is, the number of passengers that were at the stop but were not picked up by the bus because it reached . 
From these three quantities the only one that we can know with certainty is the number of passengers actually getting on. 
In the situations appearing at the left side of the plot, we observe no passenger coming on the bus and also no one is waiting at the bus stop. 
This means that in this situation the true excess demand is zero. 
Zero excess demand also can appear in situations were there are passengers coming on the bus, and no one else is waiting at the stop to board (right part of the plot). 
However, in the middle part of the plot, there is excess demand, since there are commuters waiting at the stop but either none was allowed to board because the bus was completely full, or only a number of them was allowed, until the load on the bus reach its capacity (i.e., on $<$ wait).  
Our first goal is to identify the situations that fall in this middle part of the setting, where the excess demand is larger than zero.

Apart from the number of passengers boarding at each stop, we also know the load of the bus as it arrives at the bus stop. 
We can use this to develop a mechanism for identifying the instances where excess demand is non-zero. 
In particular, when a bus arrives at a stop and the load is already at capacity when there is excess demand we expect that no passenger will be allowed to get in (R1). 

We need to emphasize here that the observation of R1 does not necessarily mean the there is excess demand. 
For example, a full bus might not pick up any passenger from the stop examined simply because there is no one waiting to board. 
The last row of Figure \ref{fig:ed_viz} overlays the load on the bus and shows how the aforementioned detection rules can lead to false positives (i.e., identify excess demand when there is none) or false negatives (i.e., missed detection of excess demand). 
The set $\excessin$ depicted corresponds to all the instances classified by the above rules as showing excess demand. 
However, as we see there can be both false positives and false negatives. 
In particular, false positives appear when the bus arrives already at capacity at a bus stop but there are not any passengers waiting for the bus. 
In this case, R1 will declare the presence of excess demand, when in reality there is not any. 
In addition, false negatives appear when the driver allows a small number of the passengers waiting to board on the bus, even though the bus is already at capacity. 
This can be either due to the bus driver not having a good estimate of the current bus load, or due to some passengers getting off the bus and hence, the driver allowing (an equal amount of) passengers to board. 

Given the absence of ground truth in the real data in terms of the presence of excess demand or no, it is impossible to estimate the fraction of false positives and false negatives from our detection. 
However, we expect that there will be relatively few instances of false positives, since these correspond to bus stops that overall have low demand for boarding even though the route itself has fairly high demand. 
This will usually appear for bus stops towards the end of a popular route. 
More importantly, given that these stops exhibit typically low demand, erroneously filtering out these instances from the training phase of the demand model will not provide significant/sizable bias. 
Furthermore, we expect false negative instances to be few as well, since the situations where the driver picks up a (small) fraction of the people waiting at the bus stop are typically fewer as compared to the scenarios where the bus driver simply does not pick up anyone waiting at the stop. 
If we want to eliminate these false negatives, we can simply classify as instances of excess demand any situation where the load on the bus is at its capacity. 
However, while this will eliminate false negatives it will come at the cost of more false positives as one might expect. 
While for a general classification problem this will lead to decline in the accuracy performance (with regards to detecting instances of excess demand), in our case we are interested in estimating the total excess demand. 
Therefore, for this downstream task we can expect the false positives and false negatives to balance each other out to a large extent. 
More specifically, the false positives will inflate our estimation of the total excess demand, while the false negatives will shrink this estimate.  


In the following section, using simulated data we will explore how these Type I and Type II errors might impact the estimation of the excess demand. 
The benefit of simulated data is that we know the ground truth and hence, we can explore several ``what-if'' scenarios in terms of classification accuracy and excess demand volume. 

\subsection{Excess Demand Simulations}
\label{sec:simulations}

For our simulations we simulate a short bus route with 6 bus stops. 
For simplicity, and without any loss of generality, this is a pickup only route, i.e., one that every passenger is getting off at the terminal stop\footnote{In fact such routes exist in the real system as well. For example, the bus route to the Pittsburgh International Airport is a pick up only route.}. 
The number of passengers boarding on each stop follows a non-homogeneous Poisson process. 
In particular, for stop $i$ the rate of the Poisson process is given by:

\begin{equation*}
\lambda_i(t) = \begin{cases}
n_i &\text{t $\in$~off-peak hours}\\
\alpha \cdot n_i &\text{t $\in$~peak hours}
\end{cases}
\end{equation*}

where $\alpha > 1$. 
By varying the duration of the peak hours we can generate a variety of scenarios with different excess demand, leading in different sizes for $\excessin$. 
The bus is also associated with a capacity that once it is reached no other passengers are allowed on board. 
Our simulated driver is also aware in {\em real-time} of the load on the bus, and hence, is able to let on board only the allowed number of passengers. 
Simply put, if when arriving at a stop there is space on the bus for $r$ passengers, but the number of commuters waiting at the stop is $k>r$, then $k-r$ passengers will be left behind (and will constitute excess demand for that stop).  

We simulate 1,000,000 trips for this bus route. 
Each trip is randomly assigned to a peak or off-peak hour based on the ratio of the peak hours in the specific simulation. 
Once we simulate the process, we explore three different ways of building a passenger demand model. 
All models will use the same set of features and the same modeling hypothesis (i.e., a Poisson regression) but they will differ in terms of the data used to train the model. 
More specifically, we will use the following training datasets:

\begin{itemize}
    \item {\bf T1: } given that we know the ground truth in our simulations, we train our model on all data except the situations where there is true excess demand
    \item {\bf T2: } using our simulated data we emulate the situation in real life. More specifically, when there is excess demand we censor the observations for the number of passengers getting on the bus. Then using our excess demand instance detection we identify $\excessin$ and train our model using all the data except $\excessin$.
    \item {\bf T3: } using our simulated data we emulate the situation in real life just as in the case of T2. However, instead of detecting $\excessin$, we simply use all the data to train the passenger demand model.  
\end{itemize}

The independent variables that we will use for all the models are the same -- and match the ones we will use in the real life scenarios described in Section \ref{sec:model_selection}. 
In particular, we use the number of passengers picked up during the previous 3 stops, as well as, an indicator variable on whether the trip is during rush hour or not. 



\begin{figure}
     \centering
    \includegraphics[scale=0.6]{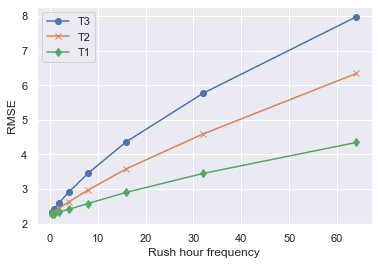}
    \caption{Removing censored observations improves the performance in terms of predicting excess demand. This improvement increases with the size of $\excessin$.}
    \label{fig:sim_filtered_VS_non}

\end{figure}

We evaluate each approach by estimating the RMSE for the (known) excess demand.  
Figure \ref{fig:sim_filtered_VS_non} depicts our results as a function of the fraction of the size of $\excessin$ relative to the total number of points simulated. 
As one might have expected the best performance in terms of RMSE is obtained when we know exactly when there was excess demand and we filter these censored observations from training. 
Nevertheless, we see that still with the imperfect detection mechanism we described in Section \ref{sec:Ie_detection} we observe significant improvement over the setting where we do not remove the censored observations. 
The benefits are larger as the size of $\excessin$ increases (relative to the total observation time). 
This is also intuitive since a larger $\excessin$ corresponds to more instances of excess demand, which if present in the training phase will increase the prediction bias. 
Delving more into the details, model {\bf T3} underpredicts the excess demand in the majority of the cases\footnote{While someone might expect this to always be the case, there are situations where this is counteracted by the large number of pickups observed the 3 previous stops. }, since the corresponding regression coefficient for the rush hour indicator variable is negative (i.e., the model trained with all the datapoints associates rush hour with less demand, which is of course contradictory by definition). 
This is of course an artifact of {\bf T3} seeing 0 passengers boarding on the bus during rush hours, but it does not know that this is simply because the bus is already at capacity. 
On the contrary, with {\bf T2} (and of course {\bf T1}), the corresponding coefficient for the rush hour indicator variable is positive, thus, correctly predicting higher volumes of (excess) demand as compared to the non-rush hours.  
These simulation results provide strong evidence that the filtering in the training phase based on the detected $\excessin$ set is able to improve the estimation of excess demand as compared to the case where these data points are not filtered out -- and hence, are considered as zero demand data points.

\subsection{Basic Data Exploration and Definitions: The 61D route}
\label{sec:61D} 

\begin{figure}
    \includegraphics[scale=0.6]{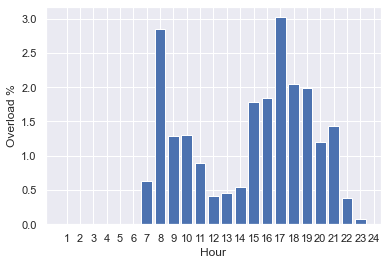}
    \caption{The probability of an overloaded bus over all stops, trips and direction for 61D.}
    \label{fig:hourly-overload}
\end{figure}

\begin{figure}
     \centering
     \begin{subfigure}[b]{0.45\textwidth}
         \centering
         \includegraphics[width=\textwidth]{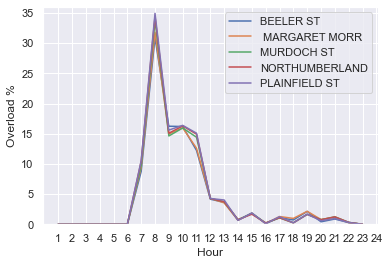}
         \caption{Inbound}
         \label{fig:inbound}
     \end{subfigure}
     \hfill
     \begin{subfigure}[b]{0.45\textwidth}
         \centering
         \includegraphics[width=\textwidth]{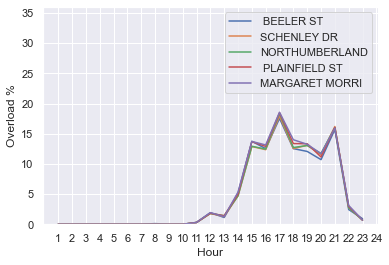}
         \caption{Outbound}
         \label{fig:Outbound}
     \end{subfigure}

        \caption{Different combinations of trip direction-stop exhibit different temporal patterns when it comes to their ``peak hours".}
        \label{fig:diff-stops-patterns}
\end{figure}

Before we move to the model selection phase, we would like to focus on a {\em testbed} bus route, namely, 61D, for understanding better the data and the environment we are dealing with. 
This route is chosen as a testbed since it goes through a diverse set of neighborhoods, including a residential area, a college/university hub and downtown Pittsburgh. 
We start by calculating the probability of observing an overloaded 61D bus. 
Figure \ref{fig:hourly-overload} presents the overall empirical probability of observing a bus with occupancy above the overload threshold aggregated across all bus stops, trips (i.e., time of the day) and trip direction  (i.e., inbound/outbound). 
We can see that overall there are two periods that emerge as ``peak hours'' for 61D overall, namely around 8 am and 5pm. 
During these times of the day there is an increased probability of a 61D rider seeing an overloaded bus regardless of the bus stop or trip direction. 
Note here that for different bus routes, the peak hours do not necessarily match those from 61D and also they do not necessarily correspond to the traditional ``rush hours'' we are accustomed to think of (morning and evening commute). 
Even for the same route, when focusing on different stops we might observe different behavior with regards to their ``peak hours'' despite the fact that in aggregate (Figure \ref{fig:hourly-overload}) these peak hours match our intuition. 
In fact, delving deeper in the results from Figure \ref{fig:hourly-overload}, the bimodal pattern observed  emerges due to the aggregation of both inbound and outbound directions, that is, inbound stops exhibit on average the early overload pattern, while the outbound stops exhibit on average the later overload pattern. 
Furthermore, aggregation across bus stops has the potential to undermine the excess demand problem in parts of the route.
In particular, during the peak hours for 61D, across all stops the maximum probability of an overloaded bus is a little higher than 3\%. 
However, looking at each stop separately reveals a different picture and points to the heterogeneity across bus stops.  
Figure \ref{fig:diff-stops-patterns} shows the probability of an overload bus arriving across 5 inbound and 5 outbound stops of 61D. 
As we can see, for the inbound route, 61D arrives overloaded approximately 35\% of the time around 8am, and in general substantial chances of an overload bus until 11am. 
For the outbound route, more than 10\% of the time 61D arrives overloaded during the evening peak hours in the same 5 stops - in fact this probability remains high for an extended period of time between 3pm-9pm, extending beyond the aggregate evening peak hours for the route (Figure \ref{fig:hourly-overload}).  
These observations make it clear that each combination of bus stop and route might exhibit different peak hour and demand patterns.
Therefore, we define the peak hour on the basis of a combination of bus stop and route as follows: 

\begin{definition}
Consider a bus stop $s$ and a route $r$. With $\pi_{sr}$ being the maximum probability of overload $P(LOAD_{sr} > Overload)$ observed for $s$ and $r$, we define its peak hour as the period of time for which, $P(LOAD_{sr} > Overload) \ge 0.75\cdot \pi_{sr}$.
\end{definition}

\begin{figure*}
    \centering
    \includegraphics[scale=0.45]{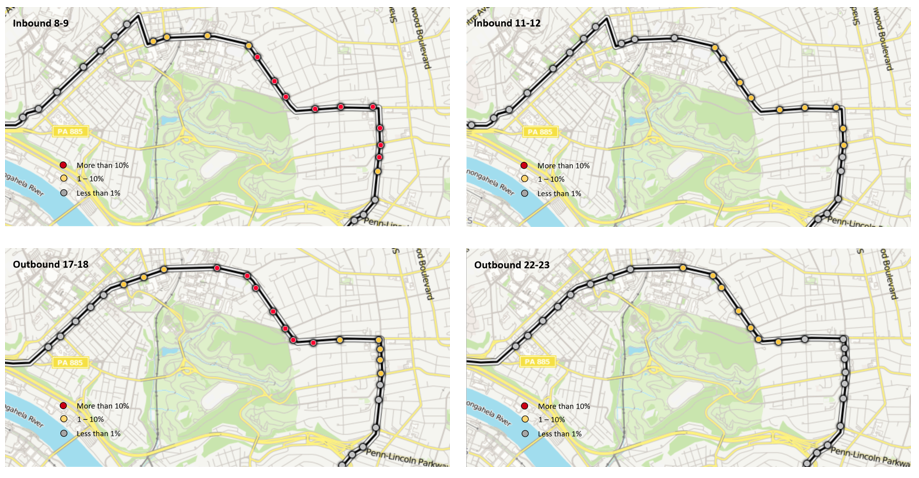}
    \caption{The probability of a bus arriving overloaded at a stop is different for different stops and times of the day.}
    \label{fig:61D-map}
\end{figure*}

\begin{figure}
    \centering
    \includegraphics[scale=0.6]{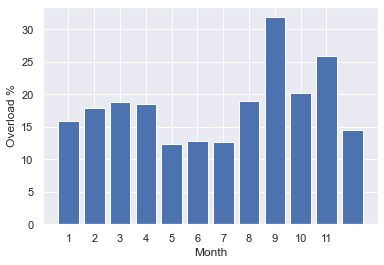}
    \caption{Probability of overload for FORBES AVE AT MARGARET MORRISON bus stop (61D) at 5pm. }
    \label{fig:forbes-margaret-monthly}
\end{figure}

To further visualize the spatio-temporal dependencies of the excess demand across the 61D route, Figure \ref{fig:61D-map} visualizes the overload probability for all the stops in the 61D inbound and outbound routes for different times of the day. 
As we can see there is a strong spatio-temporal component for the probability of excess demand at different stops, which is in alignment with the observation above that different bus stops have different peak hours. 
Furthermore, it is expected that the excess demand will be different during different times of the year. 
For example, Figure \ref{fig:forbes-margaret-monthly} shows the probability that there will be non-zero excess demand at stop ``E19520'' (Forbes ave at Margaret Morrison) of 61D at 5pm for each month of the year. 
As we can see, this probability is higher than 10\% during the whole year, with a maximum of about 32\% during September. 

\begin{table}[h]
    \begin{center}
     \begin{tabular}{|c || c |} 
     \hline
    Non-full bus & $\approx 27\%$ \\
    \hline
    passengers board \& Non-full bus & $\approx 71\%$\\
    \hline
    none boards \& Full-bus & $<2\%$ \\
     \hline
    \end{tabular}
    \caption{Data type shares averaged across top 10 bus routes based on average load}
    \label{table:data_type_shares}
    \end{center}
\end{table}

We would like to note here that when we look the PPA system as a whole the majority of the times the bus does not arrive full at the bus stop.  
In particular, Table \ref{table:data_type_shares} shows some overall statistics for the bus demand in PPA. 
Columns 1 and 2 show that on average 98\% the times the bus arrives at a stop not full, while column 3 tells us that less than 2\% of the time the bus arrives full, which could result censored observations. 
However, as aforementioned these situations are not uniformly distributed across routes, stops and times.  
For instance, for the inbound route 61D, its top-5 stops based on the average load are full approximately $21\%$ of the time during the rush hour. 
Given this (expected) dependency of the passenger demand on the time, origin (bus stop) and destination (bus route), the  dependent variable of our model is the number of passengers $Y_{ijt}$ that get on the bus route $i$ at bus station $j$, during a time period $t$\footnote{Time period $t$ here corresponds to the month of the year. I.e., for each month we train a separate model}. 
In other words, each triplet of bus route, bus stop and time period will have its own demand model.

\section{Model Selection and Results}\label{sec:model_selection}

In our simulated results we have assumed a Poisson passenger arrival process and hence, we did not have to identify the most appropriate type of distribution for modeling the data. 
Obviously, this is not the case in real data. 
In this section we will explore 4 different models for the passenger demand, and select the {\em best} through validation. 
Fast forwarding to our model selection's result we find that the Poisson model indeed provides the best results in terms of out-of-sample prediction. 
However, we should note here that for other bus systems in other cities with different characteristics a different model might be more appropriate. 
For instance, the demand in NYC's bus system might exhibit overdispersion, making a negative binomial model more appropriate. 
So replicating our approach to a different city requires going through the model selection process as well, rather than choosing blindly the Poisson model that works best for our data from PPA. 

To reiterate our dependent variable $Y_{ijt}$ represents the number of passengers boarding on the bus route $i$ at bus station $j$, during a time period $t$. 
In what follows we start by describing the different models we considered as well as our independent variables: 

\begin{enumerate}
    \item {\bf Poisson regression}: In this case, we model the average passenger demand rate $Y_{ijt}$ through a linear combination of the set of independent variables $\mathbf{X}$ as:
    
    \begin{equation}
     \lambda_{Y} = e^{\alpha + (\mathbf{b}\cdot \mathbf{X})}
     \label{eq:poisson}
    \end{equation}
    
    The parameters $\alpha$ and $\mathbf{b}$ are obtained through maximum likelihood estimation and we can thus, estimate the distribution for $Y_{ijt}$ as: 
    
    \begin{equation}
     p(Y_{ijt}= k | \mathbf{X}, \mathbf{b}, \alpha) = \dfrac{e^{k\cdot(\alpha + (\mathbf{b}\cdot \mathbf{X}))}}{k!}\cdot e^{-e^{\alpha + (\mathbf{b}\cdot \mathbf{X})}}
    \end{equation}

    \item {\bf Negative binomial regression}: This is a typical alternative to a Poisson regression model, when the dependent variable exhibits overdispersion, i.e., the variance is larger than the expected value. This is again a generalized linear model, where the dependent variable is assumed to follow a negative binomial distribution and is estimated through maximum likelihood estimation. 
    \item {\bf Zero-inflated Poisson regression}: This is an extension of the Poisson regression that deals with situations were there is an excess of 0s (as compared to what a pure Poisson model would predict). The model mixes two processes; one that generates zeros and one that follows Poisson distribution and generates counts. 
    In particular,

    \begin{equation}
    P(Y=y) = \begin{cases} \pi + (1-\pi)e^{-\lambda} & y = 0 \\
    \frac{(1-\pi)\lambda^{{y}} e^{-\lambda}}{{{y} !}} & y \in \mathbb{Z}^+ \end{cases}
    \end{equation}
   
   Parameters $\pi$ and $\lambda$ are parametrized using the model covariates. 
   While one does not necessarily needs to use the same covariates to model the two parameters, in our case the set of independent variables we use for both $\pi$ and $\lambda$ are the same.

   \item {\bf Hierarchical model}: This is an empirical model we designed for our setting that operates in two steps. 
   First, we built a logistic regression model for the probability $p_{excess}$ of the excess demand being non-zero by using the set $\excessin$ as the positive class. 
   Then, we use a Poisson regression model to estimate the average passenger demand rate $\lambda$ (for bus route $i$ at bus station $j$, during a time period $t$). 
   We finally predict the excess demand as $p_{excess}\cdot \lambda$.  

 
\end{enumerate}

Based on our results from the simulations in Section \ref{sec:simulations} we train all the aforementioned models by filtering out the data points in $\excessin$. 
Furthermore, we use the same set of independent variables for all of the models as described in the following: 

\begin{itemize}
    \item Time $\timex$: the time that the bus arrives at the stop. This is a categorial variable and we round the time at the nearest top of the hour. 
    \item Bus inter-arrival time $\businter$: the amount of time passed since the arrival of the previous bus at the stop - for the same route.
    \item Bus scheduled inter-arrival time $\busintersched$: the scheduled time between the previous and the current bus. 
    \item Local passenger load $\lastthree$: the total number of passengers that boarded on the bus during the last three stops. 
\end{itemize}

\begin{figure}[h!]
    \centering
    \includegraphics[width=0.4\textwidth]{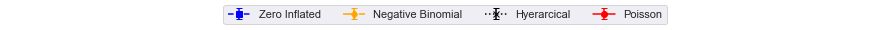}
    \centering
     \begin{subfigure}[b]{0.2\textwidth}
         \centering
         \includegraphics[width=\textwidth]{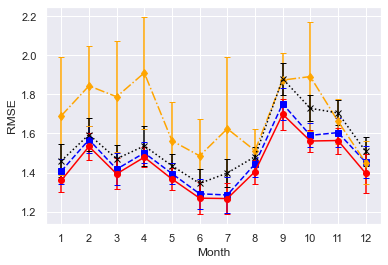}
         \caption{All bus stops}
         \label{fig:RMSE_ALL}
     \end{subfigure}
     \hfill
     \begin{subfigure}[b]{0.2\textwidth}
         \centering
         \includegraphics[width=\textwidth]{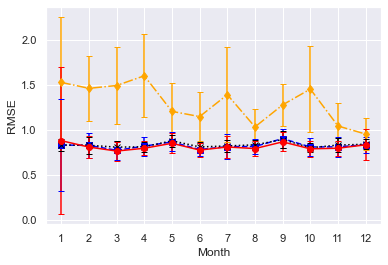}
         \caption{Low-demand bus stops}
         \label{fig:RMSE_LOW}
     \end{subfigure}
     \hfill
     \begin{subfigure}[b]{0.2\textwidth}
         \centering
         \includegraphics[width=\textwidth]{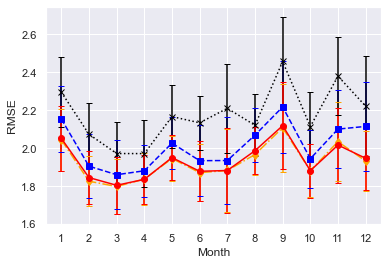}
         \caption{Medium-demand bus stops}
         \label{fig:RMSE_MEDIUM}
     \end{subfigure}
     \hfill
     \begin{subfigure}[b]{0.2\textwidth}
         \centering
         \includegraphics[width=\textwidth]{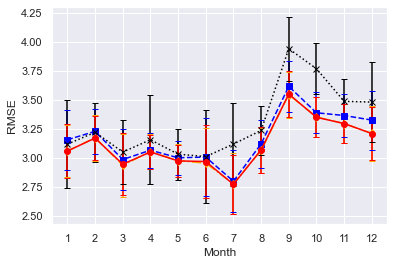}
         \caption{High-demand bus stops}
         \label{fig:RMSE_HIGH}
     \end{subfigure}

        \caption{Monthly RMSE for all, low, medium and high demanded bus stops.}
        \label{fig:614_all}
        \vspace{-0.1in}
\end{figure}

Note that we use both the scheduled inter-arrival time $\busintersched$, as well as, the actual inter-arrival time $\businter$, since (severe) delays might result in the accumulation of passengers beyond the expected based on the schedule.  
Moreover, we do not use an indicator for the yearly period, since we will build a separate model for each month of the year, i.e., using data only from the trips of the corresponding month. 
Focusing first - for illustrative purposes - to the 61D route, we train the models on 80\% of the data and calculate the Root Mean Squared Error (RMSE) of the predictions from each model on the validation/hold-out set. 
The results for all the stops in 61D are presented in Figure \ref{fig:RMSE_ALL} where the x-axis corresponds to the month of the year and the y-axis corresponds to the respective RMSE as calculated on the validation set. 
We further obtain the bootstrapped 95\% confidence intervals, using 100 bootstrap samples. 
The overall RMSE across all months and bus stops for the Poisson model is 1.44.

For robustness, we also analyze the models' performance on different types of stops based on their traffic profile. 
In particular, we define low, medium and high demand bus stops those with an average number of passengers boarding on the bus per trip less than 1, between 1 and 2, and higher than 2 respectively. 
These results are presented at Figures \ref{fig:614_all} B-D. 
As we can see, overall the Poisson regression model performs the best in the vast majority of the cases. 
Of course, based on the bootstrapped confidence intervals we can see that the differences among the various models are not necessarily statistically significant. 
Even though the performance improvements overall are not huge, we will use the Poisson model for the rest of our analysis. 

\begin{figure}[h!]
    \centering
    \includegraphics[scale=0.28]{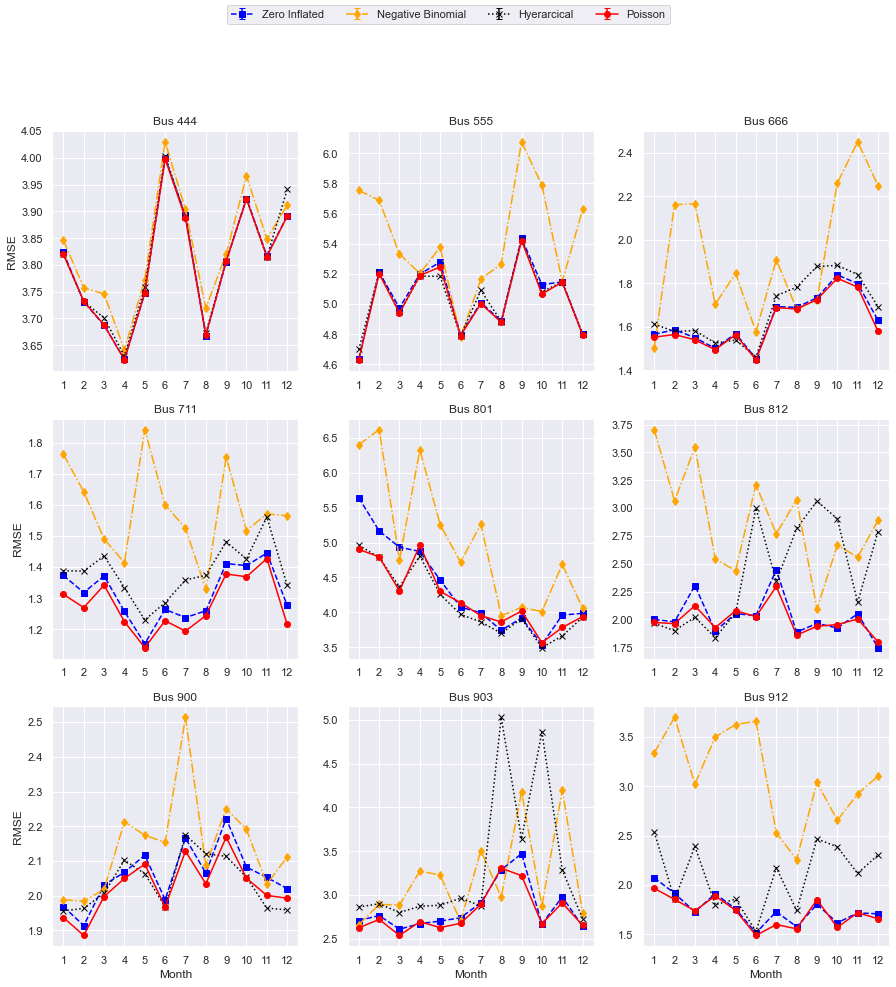}
    \caption{Monthly RMSE for various bus routes.}
    \label{fig:all_bus_RMSE}
    \vspace{-0.1in}
\end{figure}

In particular, we repeated the same process for the remaining 9 out of the top-10 routes based on average load.  
Figure \ref{fig:all_bus_RMSE} presents the RMSE for all the models across all bus stops. 
As we can see, the Poisson model still is overall the model that provides the lower out-of-sample RMSE.


Using the Poisson model learnt in the previous section we can now estimate the total excess demand over the top-10 bus routes. 
In order to estimate the total excess demand we focus on making predictions on the data points in $\excessin$. 
Recall that we filtered out these data from the training phase since these are the situations where there is potential excess demand, and to avoid biasing the models. 
This means that these predictions are out-of-sample.

\begin{figure}[h!]
    \centering
     \begin{subfigure}[b]{0.4\textwidth}
         \centering
         \includegraphics[width=\textwidth]{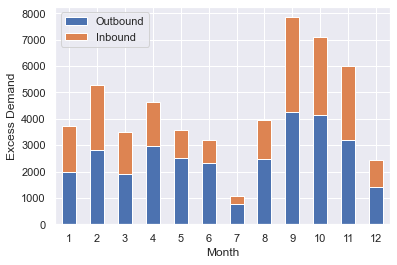}
         \caption{Absolute number}
         \label{fig:unobserved_all}
     \end{subfigure}
     \hfill
     \begin{subfigure}[b]{0.4\textwidth}
         \centering
         \includegraphics[width=\textwidth]{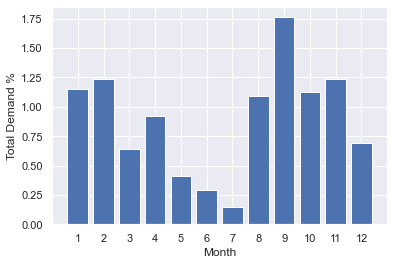}
         \caption{Fraction of passengers}
         \label{fig:unobserved_614_fraction}
     \end{subfigure}
        \caption{Total excess demand for the top-10 PPA.}
        \label{fig:total_excess}
\end{figure}

Figure \ref{fig:unobserved_all} presents our results broken down by month and route direction. 
As we can see the total number of passengers {\em left behind} is the lowest during the summer months and in December. 
This is to be expected due to the universities' recession and holidays respectively. 
The excess demand peaks again during the fall months with students coming back. 
We also present at Figure \ref{fig:unobserved_614_fraction} the excess demand estimated for the top-10 routes as a function of the number of people on the bus. 
As we can see the estimated excess demand is overall less than 2\% of the total bus load, while if we focus only on the load during the rush hours when excess demand is prevalent, this fractions increases up to 8\%. 
We leave it to the operator to make a decision on whether this number is acceptable or not, and whether they need to react as a result (e.g., with adding bus trips during excess demand etc.). 
Finally, even though this estimation is consistent with PPA running already an optimized bus schedule, it is useful for monitoring any changes in the excess demand over longer time periods.  

\section{Discussion and Conclusions}\label{sec:conclusion}

In this work we provide a framework for estimating excess demand in a public transportation system. 
Excess demand is defined as the demand that the service has, while it is at full capacity, and hence, it can neither be satisfied nor captured from traditional logging functionalities such as ticket logging. 
Our framework includes a simple detection module for detecting instances of possible excess demand based on the load of the bus arriving at a stop and the number of people boarding at that stop. 
We show through simulations that training a passenger demand model excluding these detected instances, provides a more accurate model for predicting excess demand as compared to the situation where no filtering is applied. 
The latter leads to censored observations being used in the learning process, which consequently leads to inaccuracies in predicting the passenger (excess) demand. 
We then use data from the Pittsburgh Port Authority to estimate the excess demand in the top-10 routes of the system. 
Through a model selection process a Poisson regression is chosen as the final model for the data at hand. 

While this framework is generic and can be applied in other bus systems, the results themselves might not be transferable. 
For one, the excess demands for other systems might be completely different, both in terms of relative volume as well as, in terms of its spatio-temporal characteristics. 
Furthermore, a different model could be more appropriate for a different dataset. 
An implicit assumption in our final estimation is that the excess demand follows the same distribution as the observed/measured demand. 
In particular, the model selection process is performed on a validation set from the instances where no excess demand is detected. 
However, the excess demand might for example exhibit overdispersion and hence, a negative binomial being a more appropriate model, even if Poisson is the best model for the observed demand. 
This can only be identified through field measurements, where the excess demand is explicitly measured and used for comparisons.

\small
\bibliographystyle{ACM-Reference-Format}
\bibliography{sample-base}  

\newpage

\appendix
\section{Reproducibility Appendix}

The code for our simulation described in Section \ref{sec:simulations} can be found in the following github repository: \url{https://github.com/robizon/bus_demand_estimation}. 
The bus trip data were obtained from Pittsburgh Port Authority via an NDA agreement and we are not able to provide public access. 
However, if anyone is interested we can facilitate a connection with the appropriate person at the Port Authority to explore a data sharing agreement.

\end{document}